\documentstyle[sprocl]{article}

\input{psfig.sty}

\def\Journal#1#2#3#4{{#1} {\bf #2}, #3 (#4)}

\def\NPB{{\em Nucl. Phys.} B}
\def\PLB{{\em Phys. Lett.}  B}

\def\PRD{{\em Phys. Rev.} D}

\def\NPPS{{\em Nucl. Phys. Proc. Suppl.} D}
\def\PhEs{\em Phys. Essays}
\def\CQG{\em Class. Quantum Grav.}
\def\CMP{\em Commun. Math. Phys.}
\def\PhR{\em Phys. Rev.}

\def\AIHP{\em Ann. Inst. Henry Poincar\'{e}}

\def\ra{\rightarrow}
\def\be{\begin{equation}}
\def\ee{\end{equation}}
\def\bea{\begin{eqnarray}}
\def\eea{\end{eqnarray}}

\begin{document}

\title{THE QUASI-CLASSICAL MODEL OF THE SPHERICAL CONFIGURATION
IN GENERAL RELATIVITY}

\author{VALENTIN D. GLADUSH}

\address{Department of Physics, Dnepropetrovsk University, \\
per. Nauchniy 13, Dnepropetrovsk 49050, Ukraine\\ E-mail:
gladush@ff.dsu.dp.ua}

\maketitle \abstracts {We consider the quasi-classical model of
the spin-free configuration on the basis of the self-gravitating spherical
dust shell in General Relativity. For determination of the energy spectrum
of the stationary states on the basis of quasi-classical quantization
rules it is required to carry out some regularization of the system. It is
realized by an embedding of the initial system in the extended system with
rotation. Then, the stationary states of the spherical shells are
$S$-\,states of the system with the intrinsic momentum. The
quasi-classical treatment of a stability of the configuration is
associated with the Langer modification of a square of the quantum
mechanical intrinsic momentum. It gives value of critical bare mass of the
shell determining threshold of stability.  For the shell with the bare
mass smaller or equal to the Planck's mass, the energy spectra of bound
states are found. We obtain the expression for tunneling probability of
the shell and construct the quasi-classical model of the pair creation and
annihilation of the shells.}

\section{Introduction}
Thin spherical dust shell is among the simplest popular models of
collapsing gravitating systems in General Relativity. The construction of
the canonical formalism for this model, as well as its quantization, was
considered in the works [1]-[8]. 
It was carried out in different ways and ultimately gave physically
various results. There is a number of problems here, main of which is
dependence on the choice of the evolution parameter (internal, external,
proper). The choice of time coordinate affects the choice of a particular
quantization scheme, leading, in general, to quantum theories which are
not unitarily equivalent.

At a classical level, the choice of the evolutionary parameter is the
choice of the observer, for which this parameter is the proper time. If we
choose resting (interior or exterior) or comoving observer, we obtain the
different pictures. Essential physical meaning has a picture of a
gravitational collapse from the point of view of a distant observer at
rest. In quantum theory this point of view enables us to treat bound
states in the terms of asymptotic quantities and to build the relevant
scattering theory correctly. On the other hand, the primordial black holes
in the theory of self-gravitating shells is convenient for consideration
from resting at the centre observer's point of view. In most of the works
the canonical formalism is associated with proper time of a shell. The
used reduction of the system leads to rather complicated Lagrangians and
Hamiltonians, that creates difficulties at quantization. In particular it
leads to the theories with higher derivative or to finite difference
equations.

In our opinion, the choice of the exterior or interior resting observer
is most natural. For providing of necessary properties of
invariance, the indication of canonical transformations of an extended
phase space translating the corresponding dynamic systems one in to another
is sufficient.

One of the most natural approaches to the quantization of this model it is
reduction of the problem to the usual $S$-\,wave Klein-Gordon equation in
a Coulomb field.\cite{ha5,do} However, the initial classical Hamiltonian
used here,
in fact, was postulated.

The simple and natural variational principle for a dust shell in General
Relativity can be constructed by elimination of the gravitational degrees
of freedom from approaching action containing the standard
Einstein-Hilbert term.\cite{gl} The variational formula, following from
here, leads to the required Lagrangians. They describe the shell from the
point of view of the exterior or interior resting observers. The condition
of isometric equivalence of such descriptions of the shell leads to the
momentum and Hamiltonian constraints. The corresponding dynamic systems
are canonically equivalent in the extended phase space.

In the considered model, there was a set of problems, which we would like
to discuss here. The basic problem, is why there are the stationary
quantum states for the spherical dust configuration in General Relativity,
where only an attraction takes place and there is not stable ground
nonsingular state? Why do the bound states exist at bare mass $m$ of a
shell no more than the Planck's mass only? Why is the usual quantum
mechanical consideration inapplicable at masses more than the Planck's
mass $m_{pl}$? How and why does the statement of boundary conditions near
a singularity depend from the free parameter of the theory, the bare mass
$m$? 

Let us note, that uncertainty relation is used for qualitative treatment
of this kind of problems. For example, the energy of a ground state of a
quantum oscillator can simply be obtained by means of this
relation.\cite{la1} However, for the systems under consideration having
singular, non-quadratic potentials, the straightforward application of an
uncertainty principle is difficult.

Another approach gives the quasi-classical consideration. If we introduce
a rotating shell, it is possible to note similarity of the
equations of the relativistic Kepler problem and rotating dust shell.
However, if in the first one, $L$ is the angular orbital momentum of a
particle, so in the second, $L$ is an intrinsic momentum or spin of a
shell. The mentioned similarity allows to look on the spherical quantum
collapse from the quasi-classical point of view, similarly to the Kepler
problem.

At the same time, we must keep in mind that the expression for a square
of the quantum-mechanical angular momentum $L^{2}_{q}=\hbar^2 l(l+1),\
(l=0,1,2,..)$ should be replaced by the quasi-classical expression
$L^{2}_{sc} =\hbar^2 (l+1/2)^2$ in all the quasi-classical
computations.\cite{la1,he} In the problems of the considered type such
quasi-classical replacement gives an exact asymptotic of wave function for
the radial equation following from  the Schrodinger or Klein-Gordon
equation. This replacement is justified by Langer
transformation.\cite{lan,ber}

The mentioned Langer modification can be received from the radial
Klein-Gordon equation for the relativistic dust rotating shell. The
characteristic relation connecting bare and Schwarzschild masses ($m $ and
$M $), intrinsic momentum $L$, radial adiabatic invariant $I_{R}$ and
principal quantum number $n$ (see Appendix A) follows from this equation.
Hence, it follows that the contribution in the radial adiabatic invariant
gives not the square of the quantum-mechanical intrinsic momentum
$L^{2}_{q}= \hbar^2 l(l+1)$, but its  ``renormalized'' quasi-classical
value  $ L^{2}_{sc}=L^{2}_{q}+\hbar ^2/4 $ where $\hbar ^2 /4 $ is the
Langer correction. This characteristic relation can be used as a
quasi-classical quantization rule. From the point of view of such
approach, the Bohr-Sommer\-feld rule of the orbit quantization for the
hydrogen-like atom follows from the characteristic equation and is exact
(without taking into consideration a spin of the electron).  The rule of
this kind, we shall be using to the shell as well. In addition, the
quantum states of the radially moving shell are determined as $S$-\,state
of the shell with an intrinsic momentum.

According to this treatment, for the determination of an energy spectrum
of the relativistic self-gravitating spherical shell it is necessary to
make a transition to the shell with rotation. Then, the quasi-classical
$S$-\,states ($l=0$) of such a system differ from the corresponding
classical states of the radially moving spherical shell by that there is a
``residual intrinsic momentum'' $L_{sc0}=\hbar /2$, which gives the
essential contribution to investigated processes. An effective centrifugal
potential of a repulsion which be opposed to gravitational self-action of
the shell corresponds to this momentum. Thus, in quasi-classical language,
the mechanism of a stability of the self-gravitating configuration is
formed owing to ``renormalization'' of the contribution of an intrinsic
momentum in the radial adiabatic invariant. The quantity of a ``residual
intrinsic momentum '' $L_{sc0}=\hbar /2$ assigns value of the critical
bare mass of the shell $m_{k}=m_{pl}$, determining threshold of its
stability. Notice that formal application of the quasi-classical
quantization rules to the radially moving shell (similarly to a
nonrelativistic case (see the Appendix C)) leads to the divergent
integral. Therefore, the mentioned transition to the rotating
configuration can be considered as procedure of a regularization of the
system.

Within the framework of such a quasi-classical approach, in this work one
considers the model of a configuration on the basis of the
self-gravitating spherical dust shell. We find an energy spectrum of the
bound states of such a system.
We obtain the formula of tunneling probability of the shell with negative
energy from one region of the space in another. One constructs the
quasi-classical model of the pair creation and annihilation of the shells.

The paper is organized as follows. In Sec.II we consider the dust
spherical shell in General Relativity and its Lagrange and Hamiltonian
formulation. The embedding of the considered two-dimensional dynamic
system into the extended four-dimensional dynamic system is constructing.
For this purpose we introduce the system with proper rotation and,
necessary for its description, the collective angular coordinates. We
treat a new system as some approximation to the shell with proper
rotation. Then we carried out qualitative analysis and classification of
the trajectories of the constructed system. In Sec.III we find an energy
spectrum of bound states of the extended system and shells. Then one
considers the quasi-classical treatment of a stability of the shell at
$m<m_{pl}$ and at $m=m_{pl}$ and the statement of the boundary conditions
for the quantum mechanical equation of the shell. In Sec.IV the
quasi-classical model of the shell tunneling from one region into another
of the analytically extended Schwarzschild space-time is constructed. The
formula for the tunneling probability of the shell is obtained. In Sec.V
we construct the quasi-classical model of the pair creation and
annihilation of the shells.

In the Appendix A the characteristic relation associating the radial
adiabatic invariant $I_{R} $, the parameters and conserved quantities of
the theory ($m$, $M$, $L$) with a principal quantum number $n$ has been
obtained. In the Appendix B the indeterminacy relation is used for the
obtaining of the energy of a ground state of the shell with the critical
mass $m=m_k$. In the Appendix C, for comparison with a relativistic shell,
the nonrelativistic dust spherical shell with $m\ll m_{pl}$ (very light
shell) and its Bohr quantization are briefly considered.

In this work we study relativistic and non-relativistic systems as well.
In this connection, we shall keep all the dimensional constants. Here $c$
is the speed of light, $\gamma$ is the gravitational constant, $\chi =
8\pi\gamma/c^2$, $\hbar$ is the Planck's constant. The metric tensor
$g_{\mu\nu} \ (\mu, \nu = 0,1,2,3)$ has the signature $(+\ -\ -\ -)$.

\section{Spherical dust shell in General Relativity}
\subsection{Radially moving spherical shell}

Let us consider a thin dust spherical shell with a surface density of bare
mass $\sigma $ in spherically-symmetric space-time $V^{(4)}$. The world
leaf of the shell forms a timelike hypersurface $\Sigma $, which divides
$V^{(4)} $ into exterior and interior regions $D^{(4)}_{\pm}$. Using the
curvature coordinates, we can choose general spatial spherical coordinates
$\{r,~\theta,~\varphi\}$ and individual time coordinates $t_{\pm}$ for
$D^{(4)}_{\pm}$. Then the hypersurface $\Sigma$ is given by the equation
$r=R_{-}(t_{-})$ in the interior region, and by $r=R_{+}(t_{+})$ in the exterior
one. When choosing appropriate times $t_{\pm}$ it is possible to put
$R_{-}(t_{-})=R_{+}(t_{+})$. Thus, any particle of the shell is described
by one collective dynamic coordinate $R=R_{\pm}(t_{\pm})$ and by two
fixed individual (Lagrange) angular coordinates $\theta$ and $\varphi $.

The gravitational fields in regions $D^{(4)}_{\pm}$ are set by the metrics
\be\label{3.1}
      ~^{(4)}ds^{2}_{\pm} =
    ~^{(2)}ds^{2}_{\pm} -r^2 (d\theta^2+\sin^2 \theta d\varphi^2)\, ,
\ee
where
\be\label{3.2}
       ~^{(2)}ds^{2}_{\pm}=f_{\pm}c^2 dt^{2}_{\pm}-
       f^{-1}_{\pm}dr^2\, , \quad
      f_{\pm} = 1 - \frac{2\gamma M_{\pm}}{c^2 r}\, ,
\ee
and $M_{+}$ and $M_{-}$ are Schwarzschild masses. We suppose, that $R >
2\gamma M_{\pm}/c^2$.

The effective action for a shell can be represented in the two
forms,\cite{gl}
\be\label{3.3}
   I_{\pm} = \frac{1}{2}\int{\cal L}_{\pm} dt_{|\pm}
      =-\frac{1}{2}\int \bigl(mc~^{(2)}ds \mp U_{(G)}dt\bigr)_{|\pm}\, ,
\ee
where
\be\label{3.4}
     {\cal L}_{\pm} =
      -mc^2 \sqrt{f_{\pm}- f^{-1}_{\pm} R^2_{t\pm}/c^2}
      \pm  U_{(G)}
\ee
are the effective Lagrangians of the dust shell. The subscripts $({\pm})$
designate, that the marked quantities are calculated with respect to the
exterior or interior coordinates, respectively. Further, $~m=4\pi\sigma
R^2$ is a bare mass of the shell, $~R_{t\pm}=dR/dt_{\pm}$ are velocities in
coordinate frames $\{t_{\pm}, R\}$, and
\be\label{3.5}
      U_{(G)} = - \frac{\gamma m^2}{2R}
\ee
is effective potential energy of gravitational self-action of the shell.

For systems with Lagrangians (\ref{3.4}), momenta and the Hamiltonians
have the form
\bea
      P_{\pm} = \frac{mR_{t\pm}}
      {f_{\pm} \sqrt{f_{\pm} - f^{-1}_{\pm} R^2_{t\pm}/c^2}}
      = \frac{m}{f_{\pm}}\frac{d R}{d\tau}\, , \label{3.7}
\eea
\bea
      H_{\pm} = c \sqrt{f_{\pm}(m^2 c^2 + f_{\pm}P_{\pm}^2)}
      \mp U_{(G)} = mc^2 f_{\pm}\ \frac{d t_{\pm}}{d \tau}
      \mp U_{(G)}
       \, ,    \label{3.8}
\eea
where $\tau $ is proper time of the shell.

The dynamic systems with Lagrangians ${\cal L}_{\pm}$ are not independent.
They satisfy the momentum and Hamiltonian constraints \cite{gl}
\bea
      & f_{+} P_{+} = f_{-} P_{-} \, , \label{3.9a}\\
      & H_{+}= H_{-}=(M_{+}-M_{-})c^2\, .\label{3.9}
\eea
In addition, these dynamic systems are canonically equivalent in the
extended phase space of the variables $\{P_{\pm},~H_{\pm},~R,~t_{\pm} \}$.
It is pointing out that the Lagrangians ${\cal L}_{\pm}$ describe the same
system, but from the different resting observer point of view. The
transition from the exterior observer to the interior one and vice versa
induces discrete gauge transformations
\be\label{3.10}
      M_{\pm} \leftrightarrow M_{\mp} \quad
      (f_{\pm} \leftrightarrow f_{\mp}), \qquad
      U_{(G)} \leftrightarrow - ~\ U_{(G)}, \qquad
      t_{\pm} \leftrightarrow t_{\mp}\, .
\ee

For a self-gravitating shell $M_{-}=0$. Having introduced a designation
$M_{+}=M$, we have $f_{-}=1, \ f_{+}=1-2\gamma M/c^2 R$. Then, constraints
(\ref{3.9a}), (\ref{3.9}) reads
\be\label{3.12}
    P_{-}=\left(1- \frac{2\gamma M}{c^2 R}\right)P_{+}\, ,
    \qquad  H_{-} = H_{+} = Mc^2 \, .
\ee
In this case the interchange of the exterior on interior observers induces
the transformations $M\leftrightarrow 0\
      (f_{+}\leftrightarrow 1),\
      U_{(G)}\leftrightarrow - U_{(G)},\
      t_{+}\leftrightarrow t_{-} $.
Using the canonical equivalence, we can choose that observer, from whose
position the picture looks simpler. Here, we shall use the interior
observer frame of reference, for which $~^{(2)}ds^{2}_{-}=
^{(2)}ds^{2}_{0} =c^2 dt^{2}-dR^2$. Therefore, for the action of the shell
we have
\be\label{3.3a}
   I = \frac{1}{2}\int\limits_{\gamma}{\cal L}d t
      =-\frac{1}{2}\int\limits_{\gamma}
      \bigl(m c~^{(2)}ds_{0} + U_{(G)}dt\bigr)\, .
\ee
In this case Lagrangian, Hamiltonian and momentum of the shell have the
simplest and natural form
\be\label{3.14}
   {\cal L}_{-} \equiv {\cal L} = -mc^2 \sqrt{1-{\dot R}^2/c^2} -  U_{(G)}\, ,
\ee
\be
    H_{-}\equiv H = c \sqrt{m^2 c^2 + P^2_{R}} + U_{(G)}\, ,
      \label{3.15}
\ee
\be
     P_{-}\equiv P_{R} = \frac{m\dot R}{\sqrt{1 -{\dot R}^2/c^2}}
      = m~ \frac{d R}{d\tau} \, , \label{3.16}
\ee
where we designated $t_{-}=t$, $\dot R=dR/dt$.

\subsection{Spherical dust shell with rotation}

Consider now extended variant of the initial model (\ref{3.3a}),
we need in the consequent quasi-classical quantization. Assume, that the
spherical dust shell, except for radial motions, can make also rotary
motions around the centre, as a whole, not breaking spherical symmetry
(i.e. as a rigid body) and, not perturbing the gravitational field. A new
system can be considered, as some approximation to a relativistic dust
shell with proper rotation. This approximation allows us, for the complete
description of the system, to be restricted by the collective coordinates
and, still, to consider the shell, as a dynamic system with a finite
number of degrees of freedom. The introduced system, except for the
collective radial coordinate $R=R(t)$, has still angular collective
coordinates, which can be taken to be Euler angles $\{ \theta=\theta(t), \
\psi=\psi(t), \ \varphi=\varphi(t)\} $.

For construction of the Lagrangian of the new system we shall take the
expression for the kinetic energy of a rotary motion of a spherical
top,\cite{go}
\be
        T_{rot}=\frac{I}{2} \left( {\dot{\theta}}^2
        + {\dot{\psi}}^2 + {\dot{\varphi}}^2
        + 2\dot{\varphi}\dot{\psi}\cos\theta \right)\, .   \label{3.17}
\ee
Here $I$ is a moment of inertia. Considering the shell, as a massive
spherical hollow top with the surface density of mass $\sigma=\sigma(R,t)$
and radius $R=R(t)$, we find
\be
       I = 4\pi\sigma R^4 = mR^2 \, .    \label{3.18}
\ee

From (\ref{3.17}) it can be seen, that the metric in configurational space
of rotary degrees of freedom of the shell is defined by the differential
quadratic form
\be
  dl^2=R^2 ({d\theta}^2 + {d\psi}^2 + {d\varphi}^2
        + 2{d\varphi}{d\psi}\cos\theta)    \label{3.19}
\ee
which is the metric of the homogeneous Bianchi $IX$ space.\cite{la2}

In addition, since the shell is still spherical, then the form of
gravitational self-action of the shell $U_{(G)}$ is conserved. Therefore,
the action of the new system can be obtained from the action (\ref{3.3a})
by the replacement
\be\label{3.20}
      ~^{(2)}ds^{2}_{0} \ra ~^{(5)}ds^{2}=c^2 dt^{2}- dR^2
      -R^2( {d\theta}^2 + {d\psi}^2 + {d\varphi}^2
        + 2{d\varphi}{d\psi}\cos\theta) \, .
\ee
Then the Lagrangian of the extended ``five-dimensional'' system has the
form
\be\label{3.20a}
      {\cal L} = -mc \sqrt{c^2- {\dot R}^2 -R^2({\dot \theta}^2
      + {\dot{\psi}}^2 + {\dot{\varphi}}^2
      + 2\dot{\varphi}\dot{\psi}\cos\theta)} -  U_{(G)}\, ,
\ee
The system being considered, in addition to energy, has the following
integrals of motion
\bea\label{3.21a}
  P_{\psi} &=& mc R^2 \left(\frac{d\psi}{ds}+
  \cos\theta\ \frac{d\varphi}{ds}\right) = \mbox{const}\, , \\
\quad P_{\varphi} &=& mc R^2 \left (\frac{d\varphi}{ds}+\cos\theta\
\frac{d\psi}{ds}\right) = \mbox{const}\, ,
\eea
\be\label{3.23}
     P^2_{\theta} + \frac{P^2_{\psi}+P^2_{\varphi}-2 P_{\psi}
      P_{\varphi}\cos\theta}{\sin^2 \theta} = \mbox{const}\, .
\ee
Here $P_{\psi},\ P_{\varphi},\ P_{\theta}$ are momenta, which are
conjugate to the coordinates $\theta, \ \psi, \ \varphi $. Using these
conservation laws, we can reduce dimensionality of the system. Supposing
in (\ref{3.21a}) $P_{\psi}=0$, we have $d\psi =-\cos\theta d\varphi $. In
this case the metric (\ref{3.20}) reduces to the four-dimensional one
\be\label{3.21b}
      ~^{(5)}ds^{2} \ra ~^{(4)}ds^{2}_{0}=c^2 dt^{2}- dR^2
      -R^2( {d\theta}^2 + \sin^2 \theta{d\varphi}^2) \, .
\ee
The Lagrangian and Hamiltonian of the extended system, obtained herewith
\be\label{3.21ab}
      {\cal L} = -mc \sqrt{c^2-{\dot R}^2 -R^2({\dot \theta}^2
      + \sin^2 \theta{~\dot \varphi}^2)} -  U_{(G)}\, ,
\ee
\be
      H = c \sqrt{m^2 c^2 + P^2_{R} + P^2_{\theta}/R^2
      + P^2_{\varphi}/R^2  \sin^2 \theta} + U_{(G)} = \mbox{const}
      \label{3.hamil}
\ee
describe a particle of mass $m$, moving in the space-time of special
relativity under the influence of the potential $U_{(G)}$. Here
$P_{R}=mcdR/ds $, $\ P_{\theta}= mc R^2 d\theta/ds$, $\ P_{\varphi}=mc R^2
\sin^2 \theta d\varphi/ds$ are momenta, which are conjugate coordinates
$R,\ \theta, \ \varphi $.

Using the expression $L^2 = P^2_{\theta} + P^2_{\varphi}/\sin^2 \theta
=\mbox{const}$ for complete intrinsic momentum of the system, and
(\ref{3.5}), the Hamiltonian (\ref{3.hamil}) can be rewritten in the form
\be\label{hamil2}
      H = c \sqrt{m^2 c^2 + P^2_{R} + \frac{L^2}{R^2}}
       - \frac{\gamma m^2}{2R}\, .
\ee
We suppose, that the Hamiltonian constraint (\ref{3.12}) is conserved for
the extended system as well. Therefore, using (\ref{hamil2}), it can be
rewritten in the form
\be\label{constraint}
    \left( Mc + \frac{\gamma m^2}{2c R}\right)^2 - P^2_{R}
    - \frac{L^2}{R^2} - m^2 c^2 = 0 \, .
\ee
Hence we find
\be\label{3.24}
      \biggl(\frac{dR}{cd\tau} \biggr)^2 = - V(R) \equiv
      \left(\frac{\gamma^2 m^2}{4c^2}
      - \frac{L^2}{m^2}\right)\frac{1}{c^2 R^2}
      + \frac{\gamma M}{c^2 R} + \frac{M^2}{m^2} -1 \, ,
\ee
where $V(R)\leq 0$ is the effective potential. Zeros of this function,
i.e. solutions to the equation
\be\label{3.25}
       V(R_m)=0\, ,
\ee
determine the turning points $R_m$ of the system. The formula for radial
acceleration of the system
\be\label{3.26}
      \frac{d^2 R}{d\tau^2} = -\frac{c^2}{2}\frac{dV}{dR} =
      - \frac{\gamma M}{2R^2}
      - \left(\frac{\gamma^2 m^2}{4c^2}
      - \frac{L^2}{m^2}\right)\frac{1}{R^3}
\ee
together with expression for velocity (\ref{3.24}) allows to carry out the
classification of the trajectories and to study their stability. Note,
that the component of acceleration, proportional to $1/R^2 $, is always
directed to the centre. The sign of the component of acceleration,
proportional to $1/R^3 $, depends on a relation between the potential
energy $U_{(G)}$ of the gravitational attraction and relativistic
``centrifugal energy'' of repulsion
\be\label{3.27}
      U_{(L)} =  \frac{cL}{R}\, .
\ee
At sufficiently large distances from the centre the first term in
(\ref{3.26}) dominates and we have an attraction. At small distances a
situation is different. If $\gamma m^2/2>cL$, then
\be\label{3.27a}
     |U_{(G)}|  > U_{(L)}
\ee
and the energy of a gravitational attraction is greater than energy of a
centrifugal repulsion, that causes a falling on the centre, i.e. a
gravitational collapse. When $|U_{(G)}|<U_{(L)}$, the repulsion
predominates above an attraction and the falling on centre is impossible.

The connection between the Schwarzschild and bare masses in a turning
point $R_m $ has the form
\be\label{3.28}
    M = - \frac{\gamma m^2}{2c^2 R_m}
         + \sqrt{m^2 + \frac{L^2}{c^2 R^{2}_{m}}}\, .
\ee
For a qualitative analysis of the shell dynamics it is convenient to
introduce the variable
\be\label{3.28a}
   {\cal M} ={\cal M}(m,~L,~R) \equiv - \frac{\gamma m^2}{2c^2 R}
         + \sqrt{m^2 + \frac{L^2}{c^2 R^{2}}}\, .
\ee
Then the relation $V(R)\leq 0$ leads to a condition
\be\label{3.28b}
   M \geq {\cal M}(m,~L,~R)\, ,
\ee
for the regions of classically admissible motions of the system with given
$M, ~m, ~L $. The equality $M={\cal M}$ is fulfilled in the turning points
$R_m $.

The extended system is invariant under the scale transformation
\be\label{rescaling}
  (M,\ m,\ P,\ L,\ R,\ \tau)\longrightarrow (a M,\ am,\ a P\ , a^2 L,\ a
  R,\  a\tau)\, .
\ee
Therefore, we can fix one of conserved parameters, for example $L$.

The character of the function ${\cal M}(R)$, at a given $L\neq 0$, is
determined by value of bare mass $m$, that can easily be seen from an
asymptotical behavior
\bea
    {\cal M}(R) =
  \left\{
  \begin{array}{lll}
    m-\gamma m^2/2c^2 R \, , &\ R \ra \infty\, , \\
    cm^2 R/2L+(cL-\gamma m^2/2 )/c^2 R,  &\ R \ra
    0\, .
  \end{array}
  \right.                  \label{3.29}
\eea Hence we see, that at $R\ra 0$, the function ${\cal M}={\cal M}(R)$
behaves as follows \bea
  \left\{
  \begin{array}{lll}
 {\cal M} \ra +\infty, &\mbox{if} \ c L >\gamma m^2/2\, ,\\
 {\cal M} \ra   0, &\mbox{if}  \ c L=\gamma m^2/2\, , \\
 {\cal M} \ra  - \infty, & \mbox{if} \ c L <\gamma m^2/2\,.
  \end{array}
  \right.                  \label{3.30}
\eea
Therefore the function $\cal M$ has the three types of behaviour. In the
coordinates $\{{\cal M},~R \}$, these cases are illustrated in Fig. 1.
Here, the asymptotes  ${\cal M}=\pm m$ are shown by dashed lines.
\begin{figure}[htbp]
\begin{tabbing}
a \hskip3.9cm a\=a \hskip3.9cm a\=a \hskip3.9cm a\kill

\psfig{file=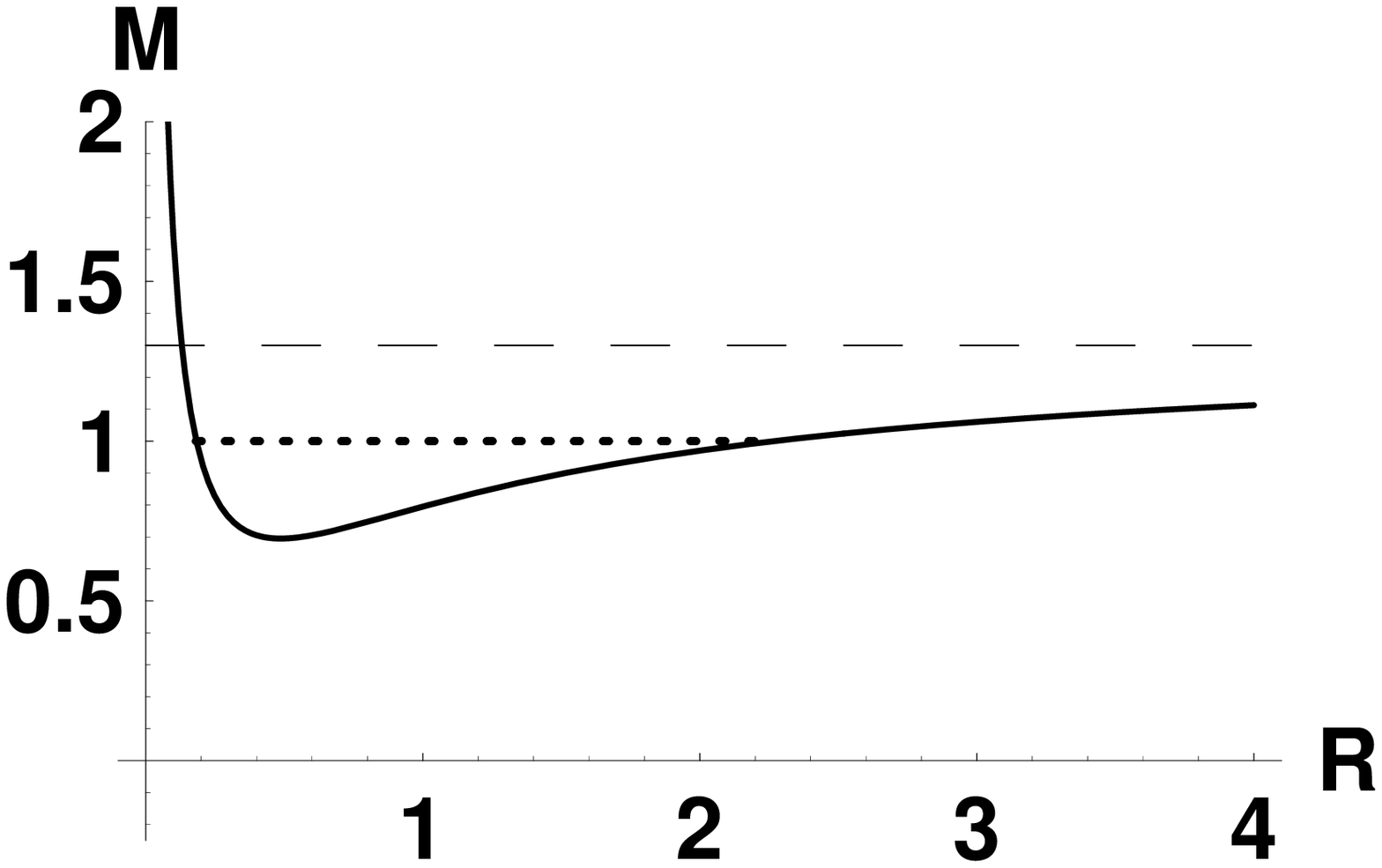,height=3.7cm,width=3.7cm}\>
\psfig{file=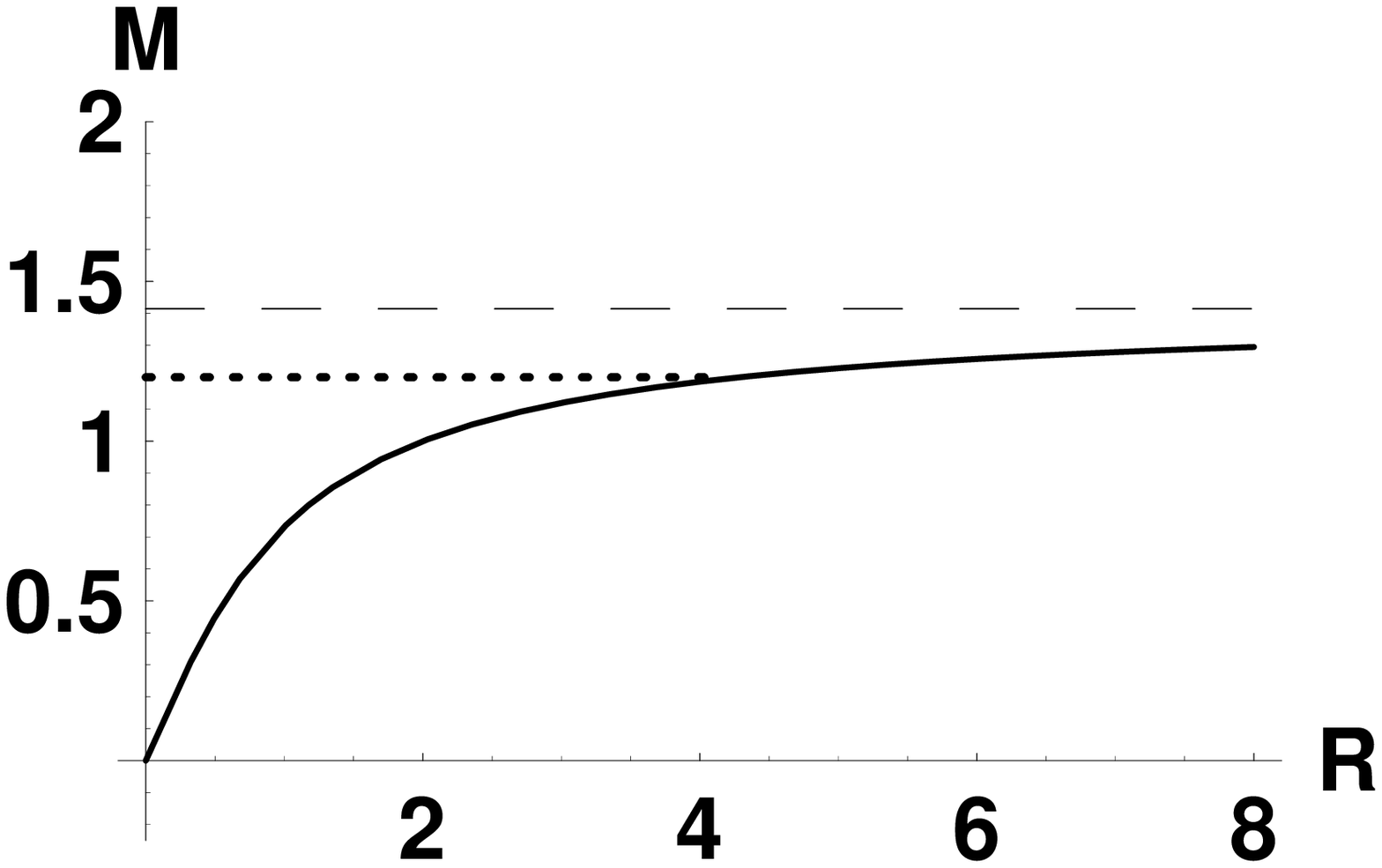,height=3.7cm,width=3.7cm}\>
\psfig{file=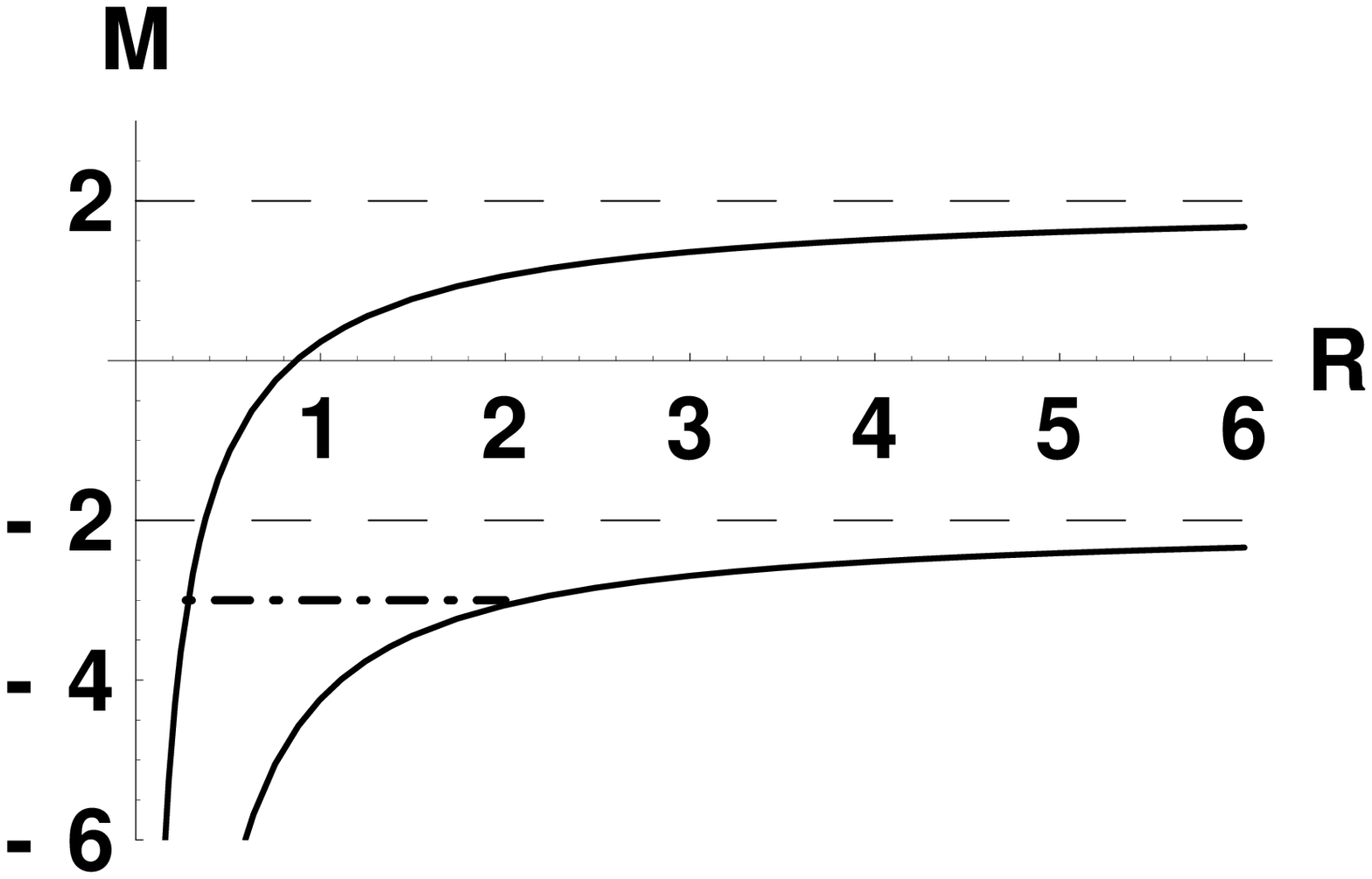,height=3.7cm,width=3.7cm}\\ \small{\qquad \qquad
\ (a)}\> \small{\qquad \qquad \ (b)}\> \small{\qquad \qquad \ (c)}\\
\end{tabbing}
\caption{The plots of the function ${\cal M = M}(R)$ for different types
of the dust shells: (a) the light shell $ 0<m< m_{k}$, (b) the shell with
the critical mass $m=m_{k}$, (c) the massive shell $m >m_{k}$. Here we
have put that $c=\gamma =L=1$, therefore $m_{k}=\sqrt{2}$.}
\end{figure}
The regions of the admissible motions are determined by segments of an
axis $R$, for which the corresponding ordinates of points of the direct
${\cal M}=M$ lay above the ordinates of points of a curve ${\cal M} ={\cal
M}(R)$ (for the lower curve in a Fig. 1(c) it is on the contrary). The
turning radii  $R _ {m} $ can be found as an abscissa of intersection
points of a curve $ {\cal M}={\cal M}(R)$ and of a direct ${\cal M}=M $.

Introduce a critical bare mass of the shell corresponding to a momentum
$L$,
\be\label{3.30a}
    m_{k}=\sqrt{\frac{2c L}{\gamma}}\, .
\ee
The relation between $m$ and $m_{k}$ determines the three possible basic
types of the shells.

\bigskip
\noindent (i) Light shell: $ 0<m< m_{k}$\,. Here $U_{(L)}>|U_{(G)|}$ and
we have an infinite potential barrier in an origin of coordinates (Fig.
1(a)). According to the value of $M$ we distinguish the variants:

\bigskip
\noindent (a) $M > m$ is a hyperbolic case. The shell with non-vanishing
initial velocity falls from infinity, reaches the turning radius
\be\label{3.31}
    R_{1}=\frac{-\gamma Mm^2 + \sqrt{\gamma^2 m^6
    +4c^2 L^2(M^2-m^2)}}{2c^2(M^2-m^2)}
\ee
and returns. The requirement $c^2 L^2 >\gamma^2 m^4/4 $ guarantees, that
$R_{1}>0$.

\bigskip
\noindent (b) $M=m$ is a parabolic case. The shell falls from the state of
rest at infinity, reaches the turning radius
\be\label{3.32}
    R_{2}=\frac{1}{\gamma m}\left( \frac{L^2}{m^2}
    -\frac{\gamma^2 m^2}{4c^2}\right)
\ee
and again returns to infinity.

\bigskip
\noindent (c) $ M_{min}\leq M<m $ is an elliptic case. The shell goes
inside of the ``potential well'' between turning radiuses
\bea\label{3.33}
    R_{3}\leq R \leq R_{4}\, ,  \qquad
    R_{3,4}=\frac{\gamma Mm^2 \pm \sqrt{\gamma^2 m^6
    -4c^2 L^2(m^2-M^2)}}
    {2c^2(m^2-M^2)}\, .
\eea
The requirement that the roots $R_{3,4}$ must be real together with the
inequality $|U_{(G)}|< U_{(L)}$ leads to the following conditions
\be\label{3.33aa}
   m^{4}_{k} \left(1-\frac{M^2}{m^2}\right) < m^4 < m^{4}_{k} \, .
\ee
The complete mass of the system $M $ can not be less than the quantity
\be\label{3.34}
     M_{min} = m\sqrt{1 - \frac{m^4}{m_{k}^{4}}}\, ,
\ee
which is the extreme of the function ${\cal M}(R)$. Therefore, we have
inequalities:
\be\label{3.34a}
      M_{min} <M <m < m_{k}\, .
\ee

\bigskip
\noindent (d) $M=M_{min}$ is a stationary case. The shell is at the bottom
of the ``potential wells''. It has minimum mass $M _ {min} $ and the fixed
radius
\be\label{3.35}
    R_{ex} = \frac{2L^2}{\gamma m^4}M_{min} =  \frac{2L^2}{\gamma
    m^3}\sqrt{1 - \frac{\gamma^2 m^4}{4c^2 L^2}}  \, .
\ee
Thus, for a light shell $U_{(L)}>|U_{(G)}|$, $M\geq M_{min}>0$, and thus,
the gravitational collapse is impossible. If bare mass reaches a critical
mass $m=m_k $, there occurs a bifurcation of the system, and the character
of trajectories varies sharply.

\bigskip
\noindent (ii) Shell with a critical mass: $m=m_{k}$\,. Here
$U_{(L)}=|U_{(G)}|$ and we have a finite potential well with the depth $m$
(Fig. 2(b)). The acceleration (\ref{3.26}) contains the component
proportional to $1/R^2 $ only. According to the value of $M$ we
distinguish the variants:

\bigskip
\noindent (a) $M>m $ is a hyperbolic case. The shell with non-vanishing
initial velocity falls from infinity to centre.

\bigskip
\noindent (b) $M=m$ is a parabolic case. The shell falls on centre from the state of
rest at infinity.

\bigskip
\noindent (c) $ 0<M<m $ is an elliptic case. The system goes in a region
\bea\label{3.36}
    0 \leq R \leq R_{5}=\frac{\gamma M  m^2}{c^2(m^2-M^2)}  \, .
\eea

\bigskip
\noindent (d)$ M=0 $ is a ``vacuum state''. The system is characterized by
the complete gravitational defect of mass, has the point sizes and rests
at the centre.

Thus, mutual compensation of the centrifugal energy and of the
gravitational energies of the self-action $(U_{(L)}=|U_{(G)}|)$ is
observed for the shell with a critical mass. It leads to the nonnegative
complete mass of the shell and to a gravitational collapse. The further
growth of bare mass leads to a new bifurcation of the system.

\bigskip
\noindent (iii) Massive shell: $ m > m_{k}$\,. Here $|U_{(G)}|>U_{(L)}$
and we have an infinite potential well (Fig. 3(c)). Depending on the value
of $M$, we have the following variants:

\bigskip
\noindent (a) $M>m $ is a hyperbolic case. The shell with nonvanishing
initial velocity falls from infinity to the centre.

\bigskip
\noindent (b) $M=m $ is a parabolic case. The shell falls to the centre
from a state of rest of infinity.

\bigskip
\noindent (c) $0\leq M<m $ is an elliptic case. The system moves within
the region
\bea\label{3.37}
    0 < R \leq R_{6} =
    \frac{\gamma Mm^2 + \sqrt{\gamma^2 m^6
    -4c^2 L^2(m^2-M^2)}}
    {2c^2(m^2-M^2)}\, .
\eea
The turning radius $R_{6}$ is real, if $~m^4_{k}(1-M^2/m^2)<m^4$.

\bigskip
\noindent (d) $M=0$ is a ``vacuum state''. The system is characterized by
the complete gravitational defect of mass. Its motion with respect to the
interior resting observer's clock, is described by the equation of an
oscillator
\be\label{oscill}
  \left(\frac{d R}{dt}\right)^2 + \omega^2 R^2 =\omega^2 R^{2}_{7}\, ,
\ee
where
\be\label{freq}
  \omega =\frac{2c^3}{\gamma m}\, , \quad
  R_{7}= \sqrt{\frac{\gamma^2 m^2}{4c^4}-\frac{L^2}{m^2 c^2}}\, .
\ee
The shell falls from a point $R_{7}$ and in time $T=\pi\gamma m/2 c^3 $
reaches the centre.

If one will prolong all the relations into the region of negative masses
$M$, we should, for generality, take into account both signs before the
radical in the expressions (\ref{3.28}) and (\ref{3.28a}). It corresponds
to the fact that the states with negative bare mass $m$ are allowed. Then
the following variants can be added:

\bigskip
\noindent (e) $-m<M<0~$ is a ``state with a negative energy'' and with
region of motions
\bea\label{3.40}
    0 < R \leq R_{8} =
    \frac{-\gamma |M|m^2 + \sqrt{\gamma^2 m^6
    -4c^2 L^2(m^2-M^2)}}
    {2c^2(m^2-M^2)}\, .
\eea

\bigskip
\noindent (f) $M\leq-|m|~$ is a ``state with a negative energy and with
the two allowed regions of motions''
\be
  \{0 < R \leq R_{9}\}\, ,\ \mbox{if} \  m > m_{k}\, , \quad
  \{R_{10}\leq R < \infty\} \, ,\ \mbox{if} \  m < - m_{k}\, , \label{3.42}
\ee
where
\be
       R_{9,10} = \frac{\gamma |M|m^2 \mp \sqrt{\gamma^2 m^6
      + 4c^2 L^2(M^2-m^2)}}{2c^2(M^2-m^2)}\, .   \label{3.43}
\ee
Thus, for the massive shell $|U_{(G)}|>U_{(L)}$, and, the gravitational
collapse is inevitable.

The character of the motion of the system from the exterior observer point
of view will be quite a different and depend on a cross arrangement of the
turning radiuses and gravitational radius $R_g =2\gamma M$. In the
considered model the position of the interior observer is convenient for
description of the primordial or eternal black holes, whereas the exterior
observer position is useful for description of a gravitational collapse.

\section{Energy spectrum of a bound states of the dust spherical shell}

Consider now quasi-classical model of the like-particle
spin-free configuration with mass $M$, using of a self-gravitating
spherical dust shell with bare mass $m$. Classical action, Lagrangian,
Hamiltonian, momenta and the constraints of this system are described by
the formulas (\ref{3.12})-(\ref{3.16}).

For determination of energy levels of stationary states of the system, the
quasi-classical approach is used. Note, that formal application of the
Bohr quantization rule of the elliptic orbits, in contrast to the
nonrelativistic case (see Appendix C), leads to the divergent integral.
Therefore, we shall perform some regularization of the system. For this
purpose we shall take advantage of a method of embedding of the given
two-dimensional dynamic system into the extended four-dimensional dynamic
system. We suppose, that effective quantum dynamics of the radially moving
shell with classical action (\ref{3.3a}) is an $S$-\,state of more general
four-dimensional quantum system with the extended classical action, in
which the angular degrees of freedom $\{\theta,\ \varphi \} $ are allowed
also. The classical Lagrangian (\ref{3.21ab}) and Hamiltonian
(\ref{3.hamil}) of such a system and the classification of the
trajectories are considered in a Sec. II.2. Hence it follows that
quasi-classical stationary states can have the light shell $m<m_{k}$, when
$U_{(L)}>|U_{(G)}|$. The bound states are possible, when $M_{min}\leq M< m
$, i.e. in the case of the elliptic trajectories $1.(c)$. The
corresponding trajectory is shown on Fig. 1(a)  by the dot line. Here the
system is inside of the ``potential wells'' and goes into the restricted
region.

We shall take an advantage of the following quantization rules,\cite{la1}
\be\label{4.1}
    I_{\varphi} = \oint P_{\varphi}\, d\varphi
    =  \int \limits^{2\pi}_{0} P_{\varphi}\, d\varphi
    = 2\pi n_{\varphi} \hbar\, ,\qquad (n_{\varphi}=0,~\pm 1,~\pm 2,...) \, ,
\ee
\be\label{4.2}
     I_{\theta} = \oint P_{\theta}\, d\theta
    = 2 \int \limits^{\theta_{4}}_{\theta_{3}} P_{\theta}\, d\theta
    = 2\pi \hbar \left(n_{\theta}+\frac{1}{2}\right)\, ,
     \qquad  (n_{\theta}=0,~1,~2,...) \, ,
\ee
\be\label{4.3}
     I_{R} =  \oint P_{R}\, d R
    = 2 \int \limits^{R_{4}}_{R_{3}} P_{R}\, dR
    = 2\pi \hbar \left(n+\frac{1}{2}\right)\, ,\qquad(n=0,~1,~2,...) \, .
\ee where $I_{R},~I_{\theta},~I_{\varphi}$ are adiabatic invariants,
and $\{R_{3},~\theta_{3}\}$ and $\{R_{4},~\theta_{4}\}$ are the
turning points coordinates (\ref{3.33}) of the  system. From
(\ref{4.1}) follows that $P_{\varphi}=n_{\varphi}\hbar $. Integral in
the relation (\ref{4.2}) gives $I_{\theta}=2\pi(L-|P_{\varphi}|)$.
Hence we obtain the quasi-classical expression for an intrinsic
momentum \be\label{4.5}
     L = L_{sc} \equiv \hbar \left(l+\frac{1}{2}\right)\, ,
     \qquad (l\equiv n_{\theta}+|n_{\varphi}|=0,1,2,..) \, .
\ee
Substituting $P_{R}$ from (\ref{constraint}) into (\ref{4.3}), one obtains
the radial adiabatic invariant
\bea\label{4.5a}
    I_R = 2 \int \limits^{R_{4}}_{R_{3}}
   \sqrt{\left(c M+\frac{\gamma m^2}{2cR}\right)^2 -
        \frac{L^2}{R^2}-m^2c^2}\ \ dR \\ \nonumber
    = \frac{\pi}{c} \left(\frac{\gamma Mm^2}{\sqrt{m^2 -M^2}}
  - \sqrt{4c^2 L^2 -\gamma^2 m^4}\right) \, .
\eea

In the Appendix A is shown, that the quantization rule (\ref{4.3}), in
which it should be put $L=L_{sc}$, is the consequence of the
characteristic relation following from the Klein-Gordon equation
(\ref{a.5}) for stationary states of the extended system and is exact.
Solving the relation (\ref{4.5a}) for $M$ and taking into account
(\ref{4.3}) and (\ref{4.5}), we obtain the energy spectrum of the extended
system
\be\label{4.6}
    \varepsilon_n =\mu \left\{1
  +\mu^4 \left[2n+1+\sqrt{(2l+1)^2 -\mu^4}~\right]^{-2}\right\}^{-1/2}\, .
\ee
Here, using the Planck's units
\bea
  m_{pl} = \sqrt{\frac{\hbar c}{\gamma}}\, ,\qquad
  l_{pl} = \sqrt{\frac{\hbar \gamma}{c^3}}\, , \label{2.7}
\eea
we come to dimensionless quantities
\be\label{4.7}
  \varepsilon_{n}=\frac{M_{n}}{ m_{pl}} \, , \qquad
  \mu = \frac{m}{ m_{pl}} \, .
\ee
The required energy spectrum of bound states of the self-gravitating
spherical dust shell, according to our rule, follows from (\ref{4.6}) for
$S$-\,states, i. e. at $l=0 $:
\be\label{4.8}
  \varepsilon_n =\mu \left\{1
   +\mu^4 \left[2n+1+\sqrt{1-\mu^4}\right]^{-2}\right\}^{-1/2}\, ,
   \qquad(n=0,~1,~2,...) \, .
\ee
that coincides with the spectrum obtained by the Hamiltonian
(\ref{3.15}).\cite{ha5}

The stability of the $S$-\,states of the shell with $m\leq m_{pl}$ can be
connected with quantum indeterminacy of its position, which follows from
the indeterminacy relation
\be\label{uncertainty}
 \overline{(\triangle P_{R})^2}\
  \overline{(\triangle R)^2}
 \geq \frac{\hbar^2}{4}\, .
\ee
Here $\overline{(\triangle Y)^2}=\overline{(Y-\overline{Y})^2}$, where
$\overline{Y}$ is a mean of the dynamic quantity $Y=\{P_{R}~
\mbox{or}~R\}$. Instead of estimate $\overline{(\triangle P_{R})^2}$ and
$\overline{(\triangle R)^2}$, we can find the manifestation of the quantum
indeterminacy (\ref{uncertainty}) having estimated its contribution in a
``renormalization'' of the adiabatic invariant $I_{R}$. The character of
the contribution follows from the characteristic equation (\ref{a.9a})
(see Appendix A). We have two kinds of the contribution in a
``renormalization'' $I_{R} $. The first is the contribution to the value
$I_{R}$, owing to $I_{R}\geq\pi\hbar$. The second is the contribution to
argument $I_{R}$, which is equal $L^{2}_{q}+\hbar^2 /4$. It determines the
effective ``renormalized'' value of a square of the intrinsic momentum
$L^2_{sc}=L^{2}_{q}+\hbar^2 /4 $, therefore $L_{sc} \geq\hbar/2 $. Owing
to this we have effective ``residual'' intrinsic momentum $L_{sc0}=\hbar/2
$ in the $S$-\,state.

Consideration of the ``residual'' intrinsic shell momentum can be treated
quasi-classically as transition from a radially moving shell to the
rotating shell, with the intrinsic momentum $L_{sc0}=\hbar/2 $. The
``residual'' intrinsic momentum is appeared in the effective centrifugal
energy (\ref{3.27}) and in repulsion, corresponding to it. The
appropriating effective critical mass, according to (\ref{2.7}) and
(\ref{3.30a}), turns out to be equal the Planck's mass $m_{k}=m_{pl}$.
Therefore the condition $|U_{(G)}| < U_{(L)}$ takes the form
\be\label{4.9}
      m<m_{pl}  \quad \mbox{or}  \quad \mu <1\, .
\ee

Thus, in quasi-classical language, the mechanism of the stability of the
self-gravi\-ta\-ting configuration is formed owing to ``renormalizations''
of the contribution of an intrinsic momentum in the radial adiabatic
invariant. The quantity $L_{sc0}=\hbar /2$ assigns value of the critical
bare mass of the shell $m_{k}=m_{pl}$, determining threshold of its
stability. For a light shell $(m<m_{pl})$ the effective centrifugal forces
of the repulsion dominate above the gravitational self-action. The
difference $U_{(L)}-|U_{(G)}|$ grows beyond all the bounds when
approaching the centre. Therefore the infinite potential barrier, which
prohibit a ``falling on centre'' even in $S$-\,state, is formed. For the
shell with the critical bare mass $(m=m_{pl})$ we have
$U_{(L)}=|U_{(G)}|$. From Fig. 1(b) it can be seen, that in this case
there is a finite potential well. In the expression for the radial
acceleration (\ref{3.26}) there remains the term proportional $1/R^2 $. In
classical case it means a falling to the centre. This trajectory is shown
on Fig. 1(b)  by the dot line. In quantum case, owing to radial
indeterminacy $I_{R}\geq \pi\hbar $ the stability of the system is
conserved. The energy spectrum of stationary states of the shell follows
from (\ref{4.8}) at $\mu =1$:
\be\label{4.8a}
  \varepsilon_n = \left\{1
   + (2n+1)^{-2}\right\}^{-1/2}\, ,
   \qquad(n=0,~1,~2,...) \, .
\ee
In classical theory the value $\varepsilon =0$ corresponds to the ground
state of such system, whereas in quantum theory the value $\varepsilon_0
=1/\sqrt{2} $ does. Since the potential well is finite here, for an
estimate of the energy of the ground state we can use the indeterminacy
relations (\ref{uncertainty}) (see Appendix B). At last, for the massive
shell $(m>m_{pl})$ we have $|U_{(G)}|>U_{(L)}$ and at the centre there is
an infinite potential well. Therefore for any shell with $m>m_{pl}$ the
``falling on the centre'' takes place. The stationary states miss here,
and the energy spectrum (\ref{4.8}) lose its meaning, that is associated
with losses of the bare mass of the shell similarly to quantum evaporation
of black holes.\cite{do}

Performing in (\ref{constraint}) replacement $P_{R}\ra -\imath\hbar d/dR $
and supposing $L=0$, we obtain the radial Klein-Gordon equation for
$S$-wave \be\label{4.11}
    \left[\frac{d^2}{dx^2}+\left(\varepsilon
    +\frac{\mu^2}{2x}\right)^2 - \mu^2 \right]\psi = 0\, ,
\ee
where we used the Planck's units (\ref{2.7}) and variable $x=R/l_{pl}$.
From our consideration follows, that the statement of the boundary
conditions at the origin for this equation depends on the quantity of the
bare mass $m$ of the shell. Thus, the light shell $(m<m_{pl})$ has an
infinite potential barrier at the origin (Fig. 1(a)). Therefore it is
necessary to use boundary conditions $\psi (0)=0$. The shell with the
critical mass $(m=m_{pl})$ has a finite potential well (Fig. 1(b)),
therefore it is sufficient to require a regularity $\psi (R)$ at the
origin. In both cases the bound states are when $\varepsilon <\mu $.

For the massive shell $(m>m_{pl})$ it is impossible to be restricted to
the one-particle approach already. Here probable tunnel transitions of the
shell in other regions of the analytically extended space-time begin to
play a role (see Sec.V).\cite{do} In this case it is necessary the
boundary conditions to formulate in the terms of the conserved Noether
currents of the corresponding field theory.\cite{ha5}

\section{Tunneling dust spherical shell}

In Sec. II.2 is noted, that under the conditions $
m^2>m^{2}_{k}$ (or $\gamma m^2/2>cL$) and $M\leq -|m|$, there are two
distinct trajectory of the shell corresponding to the different signs of
the bare mass $m$ (see case $3.(f)$ classifications). The shell can be
either into interior or exterior non-overlapping regions (\ref{3.42}) of
the space (Fig. 1(c)), which are not cause relation. The transitions
between these regions, forbidden in the classical theory, are admited in
the quantum theory. These processes are possible only for the systems with
binding energy $E_b=mc^2 -Mc^2 > 2mc^2 $. Their quantum mechanical
probability is estimated by the expression
\be\label{5.1}
      W \sim \exp \left(-\Im \frac{I_R}{\hbar} \right)
      = \exp \left(-\frac{2}{\hbar}\, \Im
     \int \limits^{R_{10}}_{R_{9}}\! P_{R}\, dR \right) \, ,
\ee
where $R_{9}$ and $R_{10}$ are turning radii (\ref{3.43}). The invariant
$I_R$ is calculated on of the classically forbidden trajectory, which
corresponds to quantum mechanical tunneling through a potential barrier.
This trajectory is shown on Fig. 1(c)  by the dot-and-dash line.

Generally, the shell with negative Schwarzschild mass $M$ is inconvenient
for interpreting in the terms of the exterior removed observer.\cite{ha5}
However, the value $M\leq 0~$ seems to be quite justified, if the shell is
contained in the configuration from the set of the spherical shells
inserted each into another.

Rewrite the above mentioned conditions as $|M|>|m|>m_{k}$,\ $M<0$ and
compare with conditions (\ref{3.34a}). Hence it can be seen, that we deal
with Euclidean analog of the case of the stationary states considered in
Sec. 3. The value $I_R $ can be obtained from the expression (\ref{4.5a})
by an analytic continuation on parameters $\{M,\ m \}$ into region
$|M|>|m|>m_{k}$,\ $M<0$. It is achieved by replacement $M\ra -|M|$.
Supposing $L=\hbar/2~$ and going to dimensionless unities (\ref{4.7}), we
have \bea\label{5.3}
    I_R  =\imath\pi\hbar \left(\frac{|\varepsilon|\mu^2}
    {\sqrt{\varepsilon^2 - \mu^2}}
  - \sqrt{\mu^4 - 1}\right)\, .
\eea
As a result, the tunneling probability is
\be\label{5.4}
      W \sim \exp \left( \pi\sqrt{\mu^4 - 1} - \frac{\pi\mu^2}
    {\sqrt{1 - \mu^2/\varepsilon^2}}\right)  \qquad
    (\varepsilon <0)\, .
\ee
This formula gives an estimate of the tunneling probability of the dust
spherical shell with negative energy from one region of the space in
another. There are two thresholds at reaching which the process is
possible here. The first threshold gives an energy level in the domain of
the ground continuum $\varepsilon<-\mu <0$. The second one concerns with
the bare mass of the shell $|\mu|\geq1$. From the formula (\ref{5.4}) we
see that at $|\varepsilon|\gg\mu $ the tunneling probability is determined
by the bare mass only
\be\label{5.5}
      W \sim \exp \left(\pi \sqrt{\mu^4 - 1} - \pi\mu^2 \right)\, .
\ee In the case $\mu\gg 1$ the tunneling of the shell is, practically, a
certain event, since $W\sim\exp (-\pi/2\mu^2)\ra 1$.

\section{The pair creation of the shells and its annihilation}

Consider the manifestation of the above described tunnel
transition in the quasi-classical model the pair creation of the spherical
shells and their annihilation. It is constructed on the basis of
self-gravitating system of two concentric dust shells.

Let $R_a,~m_a,~\tau_a $ are the radius, bare mass and proper time of the
shell with the number $a$ $(a = 1,2)$. Let $ R_0 =0, \ R_1 < R_2,\ R_3 =
\infty $. We assume that $M_{a}$ is the Schwarzschild mass defining
gravitational field $f_{a}=1-2\gamma M_{a}/c^2 r$ in the region $\{R_a
<r<R_{a+1}\}\ (a=0,1,2)$. We shall designate the coordinate time in this
regions by $t_a $. Introduce the quantities $f^{-}_{a}=1- 2\gamma
M_{a-1}/c^2 R_a$ and $f^{+}_{a}=1-2\gamma M_{a}/c^2 R_a$\ $(a=1, 2)$. Then
$P^{\pm}_{a}=m_a dR_a /f^{\pm}_{a}d\tau_a $ are the momenta of the shell
with the number $a$, and $U^{(G)}_{a}=-\gamma m^{2}_{a}/2R_{a}$ is the
energy of its gravitational self-action. The expression
\be\label{6.1}
      H^{\pm}_a =
      c\varepsilon^{\pm}_{a} \sqrt{f^{\pm}_{a}(m^2_{a} c^2
      + f^{\pm}_{a}(P^{\pm}_{a})^2)} \mp U^{(G)}_{a}
      \quad (a=1,~2)\, ,
\ee
where $\varepsilon^{\pm}_{a}=\pm 1$, determines the Hamiltonians of the
$a$-th shell.\cite{gl} They are describing the dynamics of the $a$-th
shell from the point of view of the exterior or interior resting observer
in the regions $\{R_a <r<R_{a+1}\}$ or $\{R_{a-1} <r<R_{a}\}$,
respectively. By virtue of the constraints
\be
      H^{+}_a =  H^{-}_a
      = (M_a - M_{a-1})\, c^2  \quad (a=1,~2)\, ,  \label{6.2}
\ee
the total Hamiltonian of the configuration satisfies to the relation
\be\label{6.3}
      H=  H^{\pm}_{1} + H^{\pm}_{2} =(M_2 - M_0)\,c^2 \, .
\ee
The binding energy of the system is determined by the expression $E_b
=(m_1 +m_2 +M_0 -M_2)\,c^2 $.

For the self-gravitating configuration we have $M_0 =0$. In addition, we
assume that $M_2 =0$, therefore $E_b=(m_1 +m_2)c^2$. Then the Hamiltonian
constraints (\ref{6.2}) read $H^{\pm}_1=M_1 c^2 $ and $ H^{\pm}_2=- M_1
c^2 $, and the total Hamiltonian of the configuration vanishes $H=0$.
Further we are supposing that the Hamiltonian and bare mass of an exterior
shell have positive values. Then $ M_1 <0 $ and it is natural to introduce
the designation $ M_1 =-M $. From the Hamiltonians (\ref{6.1}) and
constraints (\ref{6.2}) it follows, that $ \varepsilon^{+}_{1}=-1, \
\varepsilon^{-}_{2}= 1 $. Taking into account, that
$\varepsilon^{\pm}_{a}=1 $ corresponds to the increase of the radial
coordinate in the direction of an exterior normal to the shell, it is easy
to see, that $\varepsilon^{+}_{2}=1$. The sign $\varepsilon^{-}_{1}$
remains uncertain, that corresponds to the different signs bare masses of
the interior shell $\mu^{-}_{1}=\varepsilon^{-}_{1}m_{1}$.

Thus, the gravitational field of the configuration is described by the
metric
\bea\label{6.5}
      ~^{(4)}ds^{2}=fc^2 dt^{2}- f^{-1}dr^2
      -r^2 (d\theta^2+\sin^2 \theta d\alpha^2)\, ,
\eea
where $f=1+{2\gamma M}/{c^2 r}$, if $\{R_1 <r< R_2\}$, and $f=1$, if $\{0
<r< R_1\}$ or $\{R_2 <r<\infty \}$.

Note, that each shell can be considered independently. In addition, taking
into account canonical equivalence of description of the $a$-th shell with
respect to the interior or exterior observers, it is possible to be
restricted by the simplest from the pictures and, respectively, the
simplest from the Hamiltonians $H^{\pm}_a $. Therefore we shall consider
the Hamiltonians
\be\label{6.8}
      H^{-}_1 =
      \varepsilon^{-}_{1}m_{1}c^2 \sqrt{1+\left(\frac{d R_1}{c
      d\tau_1}\right)^2} - \frac{\gamma m^{2}_{1}}{2R_{1}}
      = - M c^2 \, ,
\ee
\be\label{6.9}
      H^{+}_2 =
      m_{2}c^2 \sqrt{1+\left(\frac{d R_2}{c
      d\tau_2}\right)^2} + \frac{\gamma m^{2}_{2}}{2R_{2}}
      =  M c^2 \, ,
\ee
that corresponds to the choice of the interior observer for the first
shell and of the exterior observer for the second one. In the turning
points $R_{m 1},\ R_{m 2}$ of the shells we have
\bea\label{6.10}
    M= -\varepsilon^{-}_{1}m_{1} +\frac{\gamma m^{2}_{1}}{2R_{m1}c^2}
       = m_{2}+\frac{\gamma m^{2}_{2}}{2R_{m2}c^2} \, .
\eea
Hence it follows, that if $\varepsilon^{-}_{1}=1$, then $R_{2}>R_{1}$ at
$m_{2}\geq m_{1}$. In the case of $\varepsilon^{-}_{1}=-1$, if $
R_{m2}>R_{m1}$, then $m_{2}>m_{1}$, and if $ R_{m2}=R_{m1}$, then $
m_{2}=m_{1}=m$. In the latter case the trajectories of the shells
coincide. But their bare masses are equal in magnitude but opposite in
sign, therefore the total bare mass vanishes, i.e. shells disappear.

In the previous section we considered the process, in which the shell with
negative energy and in the state with $\mu^{-}_{1}=m$ as a result of the
tunnel transition turned out in the state with $\mu^{-}_{1}=-m$. In
accordance with that we shall consider the following situation. Let there
is the self-gravitating system of two spherical dust shells with the equal
bare mass $\mu^{-}_{1}=\mu^{+}_{2}=m$ and with equal in magnitude but
opposite in sign energies. The shells follow the trajectories $R_a =
R_a(\tau_{a})\ (a=1,2)$ and their total energy is equal to zero $E=0$. The
interior shell, as a result of the tunnel transition, jumps at the
trajectory of the exterior shell, but in the state with negative bare mass
$\mu^{-}_{1}=-m$. Thus an annihilation of the shells take place. This
process is possible at $m>m_{pl}$ and $M>m$ and its probability is
described by the formula (\ref{5.4}).

Naturally, the opposite event, i.e. the process of pair creation of
shells, is possible as well. Imagine the following picture in the flat
space-time: let two shells with equal in magnitude but opposite in sign
bare masses and energies move, as a single object, along the coincident
trajectories, which are determined by the Hamiltonians (\ref{6.8}),
(\ref{6.9}). In the turning point this vacuum system breaks up into two
distinct shells. The shell with positive bare mass and energy prolongs a
motion, and the shell with negative bare mass and energies, as a result of
the tunnel transition, jumps at the interior trajectory of the shell with
negative energy, but with positive bare mass. As a result, of it there is
the pair creation of the shells.

\section*{Acknowledgments}
I would like to thank V.~Skalozub, S.~Stepanov and A.~Skalozub for helpful
discussions of question touched in this paper.

\section*{Appendix A.  Characteristic equation}

To find connection between the radial adiabatic invariant $I_R
$, quasi-classical and quantum mechanical intrinsic momenta, we shall
consider the radial Klein-Gordon equation for the extended system. It can
be received from the Hamiltonian constraints (\ref{constraint}) by
replacement
\be\label{a.3}
    P\ra \hat{P}=-\imath\hbar\frac{\partial}{\partial R}\, ,
    \qquad
   L^2 \ra {L}^2_{q}= \hbar^2 l(l+1) \quad (l=0,1,2,...)\, .
\ee
We shall write out the equation for the radial wave function
$\psi=\psi(R)$, using the Planck's units (\ref{2.7}), dimensionless
quantities (\ref{4.7}) and variable $x=R/l_{pl}$:
\be\label{a.5}
    \left[\frac{d^{~2}}{dx^2}+\left(\varepsilon
    +\frac{\mu^2}{2x}\right)^2 - \frac{\Lambda^{2}_{q}}{x^2}
    - \mu^2 \right]\psi = 0\, ,
\ee
where $\Lambda^{2}_{q}\equiv L^{2}_{q}/\hbar^2 = l(l+1)$.

The bound states are possible, when $\gamma m^2\leq 2cL$ and $M<m $. We
impose the boundary conditions $\psi (0)=0$ and usual requirement of
vanishing of a wave function $\psi$ at infinity. As a result, the solution
of the equation (\ref{a.5}) is searched in the form
\be\label{a.7}
   \psi = z^{\frac{1+S}{2}} \exp\left(-\frac{z}{2}\right) f(z)  \, ,
\ee
where  $f(z)$ is some polynomial, and $z=2x\sqrt{\mu^2-\varepsilon^2}$. We
choose $s$, so that
\be\label{a.8a}
      s=2\sqrt{\Lambda^2_{sc}-\frac{\mu^4}{4}}\, .
\ee
Here
\be\label{a.10}
    \Lambda^2_{sc} \equiv\frac{L^{2}_{sc}}{\hbar^2}
    =\Lambda^2_{q}+\frac{1}{4}\, ,
\ee
where, by definition, quantity
\be\label{a.11}
    L^{2}_{sc}\equiv \hbar^2\Lambda^2_{sc}
    = \hbar^2{\left(l+\frac{1}{2}\right)}^2
\ee
is the square of the quasi-classical intrinsic momentum. In this case,
for $f(z)$ we obtain the equation in the canonical form
\be\label{a.9}
    z f''+(1+s-z)f'+
    \left(\frac{I_{R}}{2\pi\hbar} -\frac{1}{2}\right)f = 0\, ,
\ee
where
\be\label{a.9a}
 I_{R}
   = \pi\hbar \left(\frac{\varepsilon\mu^2}{\sqrt{\mu^2-\varepsilon^2}}
   - \sqrt{4\Lambda^2_{sc}- \mu^4} \ \right)
\ee
is the radial adiabatic invariant (\ref{4.5a}) rewritten in the Planck's
units.

If the characteristic relation is fulfilled
\be\label{a.12}
  \frac{I_{R}}{2\pi\hbar} - \frac{1}{2}=n \qquad (n=0,1,2,...),
\ee
then the solutions of the equation (\ref{a.9}) are the Laguerre
polynomials $L_{n}^{s}$.\cite{han} Therefore
\be\label{a.13}
  \psi = z^{\frac{1+S}{2}} e^{-\frac{z}{2}} L_{n}^{s}(z)\, ,
\ee
and the energy spectrum (\ref{4.6}) follows from the condition
(\ref{a.12}).

From (\ref{a.9a}) and (\ref{a.10}) it can be seen, that the argument of
the radial adiabatic invariant $I_{R}$ contains not the square of the
intrinsic momentum $L^{2}_{q}$, but the sum $ L^{2}_{sc}=L^{2}_{q}+\hbar^2
/4$. Thus, the quantum mechanical characteristic relation (\ref{a.12}),
associating both conserved quantity and quantum number $n$, is equivalent
to the quasi-classical conditions of the quantization (\ref{4.3}) and
(\ref{4.2}).

\section*{Appendix B. The shell with the critical mass and uncertainty
relation}

For a shell with the critical bare mass $(m=m_k)$ there occurs
almost complete compensation of a force of the gravitational self-action
by the effective centrifugal force of the repulsion. Therefore the finite
potential well of depth $m_k $ at $R=0 $ is formed and any values of $R$
are admited. Therefore, for an estimate of the ground state energy it is
possible to use an uncertainty principle (\ref{uncertainty}). In
$S$-\,state we have $l=0$,  $L_{sc}=L_{sc0}=\hbar/2$, $m_k= m_{pl}$. Then
the relation (\ref{hamil2}) taking into account (\ref{3.12}) and
(\ref{4.7}) can be rewritten in the form
\be\label{mass1}
 \varepsilon = \sqrt{1 + \frac{P^2_{R}}{c^2 m_{pl}^{2}} + \frac{1}{4 x^2}}
       - \frac{1}{2x}\, .
\ee
Further, we suppose that $\overline{P_{R}}=0$ and $\overline{R}=0$.
Therefore $\overline{(\triangle P_{R})^2}=\overline{P^2_{R}}$,\ \
$\overline{(\triangle R)^2}=\overline{R^2}$ and we have
\be\label{b3}
     \overline{P^2_{R}} \geq
     \frac{\hbar^2}{4l_{pl}^{2}\overline{x^2}}\, .
\ee
Assuming, that (\ref{mass1}) is satisfied, and for the mean-square
quantities we have
\be\label{mass2}
 \varepsilon \geq  \sqrt{1 + \frac{1}{2z^2}}
 - \frac{1}{2z}\, ,
\ee where $z=\sqrt{ (\overline{x^2})}$. The minimum of the energy
$\varepsilon_0=1/\sqrt{2}$ is reached at $z=1/\sqrt {2}$. This
coincides with the energy of the critical shell (\ref{4.8a}) in the
ground state at $n=0$.

\section*{Appendix C. Very light spherical shell}

For a very light shell the nonrelativistic approach is
applicable. In this case the action for a thin dust shell moving in a
proper gravitational field has the form:\cite{gl}
\bea
    I_{N} = \int \limits_{t_1}^{t_2} {\cal L}_{N}~dt
             = \int \limits_{t_1}^{t_2}\biggl(\frac{1}{2} m{\dot R}^2
             + \frac{\gamma m^2}{2R}\biggl)~dt \, . \label{2.1}
\eea
Here $m=4\pi\sigma R^2 $, $R=R (t) $  is the mass and radius of the
spherical shell, $ \dot R=dR/dt $, $ \sigma = \sigma (t) $. Hence, for
momentum and Hamiltonian of a shell, we find
\bea\label{2.2}
          P=m \dot R\, , \qquad
        H = \frac{P^2}{2m} - \frac{\gamma m^2}{2R} = E = \mbox{const}\, .
\eea
With the help of the conservation law of energy, the requirement
of applicability of nonrelativistic approach $\dot R\ll c$ can be
written as a requirement to the size of the shell $R \gg R_c =\gamma
m^2 /(m c^2-2E)$. Here $R_c $ is critical radius of a shell, at which
its velocity reaches velocity of light $c$. For hyperbolic
trajectories ($E>0 $) this radius can be somehow large and at $E=m
c^2/2$ becomes infinite. For parabolic trajectories ($E=0$) we have
$R_c =\gamma m/c^2 $. For elliptic trajectories ($E<0$) the system is
localized inside the region $0<R\leq R_{\mbox{max}}=\gamma m^2/2|E|$,
where $R_{\mbox{max}}$ is a turning point of the system.

The critical values of parameters of a shell $E $, $P $, $R\ (R\sim
R_{\mbox{max}})$ and $T$ ($T \sim $ of time of a collapse), at which
the Newtonian approach becomes inapplicable, are equal to quantum
nonrelativistic units of energy, momentum, length and time \bea
E_\gamma = \frac{\gamma^2 m^5}{\hbar^2}\, ,\qquad P_\gamma =
\frac{\gamma m^3}{\hbar}\, ,\qquad L_\gamma = \frac{\hbar^2}{\gamma
m^3}\, ,\qquad T_\gamma = \frac{\hbar^3}{\gamma^2 m^5}
\label{2.6} \eea for a particle of mass $m$ in a gravitational field
$\varphi=-\gamma m/R$. The nonrelativistic quantum approach is
possible, if $|E|\sim E_\gamma \ll mc^2 $. Hence it follows, that
this is possible for a very light shell, when $m \ll m_{pl}$
$(R_{\mbox{max}}\gg l_{pl}) $.

In order to find an energy spectrum of the bound states of the
nonrelativistic dust spherical shell with $E<0$ it is sufficient to be
restricted by the Bohr quantization rule of elliptic orbits. Since the
shell moves radially, it is possible to consider its orbit as an limiting
case of an elliptic orbit with two turning points. The first turning point
is $R_{\mbox{max}}$, as a second point is $R=0$. Indeed, the expression
(\ref{2.1}) can be treated as action for a particle of mass $m$, moving in
a central gravitational field $\varphi =-\gamma m/R $. In the
nonrelativistic case, at any non-vanishing orbital momentum, always, there
is the second turning point $R_{\mbox{min}} $ ($R_{\mbox{min}}\leq R\leq
R_{\mbox{max}}$). In the considered approximation the turning point, which
is the nearest to the centre and concerned with ``residual'' to the proper
momentum $L_{sc0}=\hbar/2 $, possibly to locate in the centre. Then the
elliptic orbit of a particle is drawn out in a line segment connecting
points $R=0$ and $R_{\mbox {max}}$.

The application of the Bohr quantization rule to radial trajectories gives
\bea\label{borh}
      \oint P\, d r
      = 2 \int \limits^{R_{\mbox{max}}}_{0}
      \sqrt{2m\left(E+\frac{\gamma m^2}{2R}\right)}\, d R
      = \frac{\pi \gamma m^{5/2}}{\sqrt{-2E}} = 2\pi n\hbar \, .
\eea
Hence, for the required spectrum of energies, we obtain
\be\label{2.9}
      E_n = - \frac{E_\gamma}{8 n^2} \qquad (n=1,2,3,...)\, .
\ee

For the very light shell $(m\ll m_{pl})$, by virtue of degeneration,
similar to the nonrelativistic Kepler problem,\cite{go} both of the above
mentioned contributions of quantum indeterminacy are added (so, that $
(n+1/2)+1/2 =n+1 $). Therefore the expression (\ref{4.8}) for the energy
spectrum transfers into expression (\ref{2.9}), in which the reading of a
quantum number $n$ begins from unity rather than from zero. Here
gravitational collapse is impossible, as the effective centrifugal energy
always is greater than energy of gravitational self-action of the shell.

\end{document}